\documentclass{cjpsuppl}
\usepackage{graphicx}

\def\x               {\chi}

\def\L               {{\cal L}}

\def\O               {{\cal O}}

\def\ti              {\tilde}

\def\nt              {\ti\x^0}
\def\ch              {\ti\x^\pm}
\def\sf              {\ti f}

\newcommand{\mh}[1] {m_{H^{#1}}}

\def\fb              {${\rm fb}^{-1}$}

\def\to              {\rightarrow}

\def\greaterthansquiggle{\raise.3ex\hbox{$>$\kern-.75em\lower1ex\hbox{$\sim$}}}
\def\lessthansquiggle{\raise.3ex\hbox{$<$\kern-.75em\lower1ex\hbox{$\sim$}}}
\def\gl{\raise.3ex\hbox{$<$\kern-.68em\lower1ex\hbox{$>$}}}

\newcommand{\lts}{\lessthansquiggle}
\newcommand{\cp}{\mbox{$\not \hspace{-0.15cm} C\!\!P \hspace{0.1cm}$}}

\def \etm{E{\!\!\!/}_T}
\def\half{\small \frac{1}{2}}

\begin{document}
\title{ CP violation in Supesymmetry and the LHC}
\authori{Rohini M. Godbole}
\addressi{Centre for High Energy Physics, Indian Institute of Science, 
Bangalore, 560 012, India.}
\authorii{}    \addressii{}
\authoriii{}   \addressiii{}
\authoriv{}    \addressiv{}
\authorv{}     \addressv{}
\authorvi{}    \addressvi{}
\headtitle{CP violation in Supersymmetry  \ldots}
\headauthor{Rohini M. Godbole}
\lastevenhead{Rohini M. Godbole CP violation in Supersymmetry 
\ldots}
\refnum{}
\daterec{ \\final version } 
\suppl{A}  \year{2005} \setcounter{page}{1}
\begin{flushright}
                                                   IISc-CHEP/3/05\\
                                                   hep-ph/0503088\\
\end{flushright}
\vspace{0.5cm}
\begin{center}
{\Large
{\bf  CP violation in Supesymmetry and the LHC }}\\[5ex]
R.M. Godbole$^a$\\[3ex]
{\it Centre for High Energy Physics, Indian Institute of Science,
Bangalore, 560 012, India.}\\[5ex]       

\end{center}
\vspace{1.0cm}
{\begin{center}
ABSTRACT
                                                                                
\vspace{2cm}
                                                                                
\parbox{15cm}{
In this talk I discuss possibilities of probing the CP violation (CPV)
in the Minimal Supersymmetric Standard Model (MSSM), at the LHC as well as
its effects  on  the LHC SUSY phenomenology. In the latter case I mainly
discuss its effect on the Higgs-sector and hence on Higgs Phenomenology at
the LHC.
After outlining the possibilities that a study of the $\ch_1 \nt_2$
production at the LHC might offer, I will  summarise the effects of
the CPV in MSSM on the Higgs searches at the  LHC. Further,
I will discuss how a study of the process $H^\pm \rightarrow W^+ \phi_1$
may be able to plug a 'hole" in the $\tan \beta$-$\mh{+}$ plane,
where the LEP has no sensitivity and where the searches in the usual discovery
channels at the LHC are likely to fail as well.}
\end{center}}
\vfil
${ }^{a}$ Talk presented at 'Physics at LHC', Vienna, 13-17 July 2004.
\newpage

\maketitle
\begin{abstract}
In this talk I discuss possibilities of probing the CP violation (CPV) 
in the Minimal Supersymmetric Standard Model (MSSM), at the LHC as well as
its effects  on  the LHC SUSY phenomenology. In the latter case I mainly 
discuss its effect on the Higgs-sector and hence on Higgs Phenomenology at
the LHC.
After outlining the possibilities that a study of the $\ch_1 \nt_2$
production at the LHC might offer, I will  summarise the effects of 
the CPV in MSSM on the Higgs searches at the  LHC. Further, 
I will discuss how a study of the process $H^\pm \rightarrow W^+ \phi_1$ 
may be able to plug a 'hole" in the $\tan \beta$-$\mh{+}$ plane, 
where the LEP has no sensitivity and where the searches in the usual discovery
channels at the LHC are likely to fail as well.
\end{abstract}
\section{Introduction}
At present Particle Physics finds itself at a very interesting juncture.
Almost all the experimental observations are explained, to a great precision
of $\sim 1$ part per mill or more, in terms of the Standard Model (SM).
The triumph of the gauge paradigm, in describing correctly the fundamental 
particles and interactions among them, is almost complete with the 2004 Nobel
Prize being awarded for the discovery of Asymptotic Freedom.  The Higgs boson
still eludes direct experimental observation, but the current precision data
and direct searches bound its mass (in the SM) in a range which is accessible 
perhaps to the Tevatron and definitely to the LHC.  In spite of this tremendous
success, the SM still does not give us a fundamental understanding of quite a
few of its features. CP violation (CPV) in the SM happens to be one of them. 
The precision measurements by BABAR and BELLE at the 
$B$--factories~\cite{bellebabar} show conclusively that {\it all} 
the CP violation observed experimentally so far, can be accurately described 
in terms of that in the up-quark mass matrix  encoded in the  phase in the 
Cabbibo-Kobayashi-Masakawa(CKM) quark-mixing matrix.  However, this amount
of CPV is not sufficient to provide a {\it quantitative} understanding of the 
observed Baryon Asymmetry(BA) i.e., 
${{N_b}\over{N_{\gamma}}} \sim 6.1 \times 10^{-10}$  {\it but}
${{N_{\bar b}}\over{N_{\gamma}}} \sim 0$. This makes a source of CPV, beyond 
that in the SM, imperative. Hence it seems logical to investigate implications
of such additional CPV for the various theoretical attempts that the Particle
Physics community 
is investigating to go beyond the SM in order to cure its various deficiencies.

Supersymmetry(SUSY), by now almost the 'standard' Beyond the Standard 
Model (BSM) physics, is arguably the best option to stabilise the Higgs mass 
(and hence the EW scale) against radiative corrections~\cite{book}. 
SUSY phenomenology at the LHC occupies a place of pride in the LHC studies, 
next only to the Higgs physics. A discussion of CPV on SUSY phenomenology 
at the LHC therefore, forms a very important part of the studies. In the 
following I will first discuss general issues about CPV and the MSSM. 
Then I will discuss the effects of CPV in SUSY on MSSM phenomenology at the 
colliders and summarise the effects of
the CPV in MSSM on the Higgs searches at the LEP and LHC. Further,
I will discuss how a study of the process $H^\pm \rightarrow W^+ \phi_1$
may be able to plug a 'hole" in the $\tan \beta$-$\mh{+}$ plane,
where the LEP has no sensitivity and where the searches in the usual discovery
channels at the LHC are likely to fail as well
\section{MSSM and CPV.}
CPV in SUSY has almost changed from an ugly duckling to a swan in the recent 
years. The most general supersymmetric version of the SM, a MSSM with complex 
SUSY parameters, contains 44 phases which can not be rotated away by a simple
redefinition of the fields. In early days of SUSY, these phases were all tuned
to zero so as to avoid unacceptably large electric dipole moments (EDM's) for 
fermions. However, a few years back it was noted~\cite{nath} that it was 
possible to satisfy all the constraints  on the EDM's with some 
of the phases of $~\O (1)$, quite generally, provided the first two 
generations of squarks are heavy.

At this point it is also worth noting that the CPV in the Higgs sector is a
very attractive source for the above mentioned additional CPV that is required
for a quantitative explanation of the observed BA in the Universe.  The 
\cp\ phases of the SUSY(breaking) parameters can induce, through loop
effects, \cp\ in a Higgs sector which is CP-conserving at the tree level.
Thus in the MSSM it maybe possible to satisfy all the EDM constraints and still 
have sufficient \cp in the theory to explain the BA quantitatively. Thus it 
is clear that \cp\ SUSY will also have implications for the Higgs Phenomenology
at the colliders. Given the fact that Higgs searches is 'raison  d'e\^tre' for 
the current and future Colliders, investigations of \cp\ in supersymmetric
theories are phenomenologically very interesting indeed.
\section{Phenomenology of the MSSM with CPV at the colliders.}
\subsection{General Remarks.}
The  independent phases in the \cp MSSM that can be large 
(up to $\sim \O (1)$), even after imposing  the EDM constraints, are
the phase of the higssino mass term $\mu$, the trilinear coupling $A_f$
as well as the gaugino masses $M_i, i = 1,2$. In addition to this, the sfermion
mass matrix also can have nonzero phases for each generation. 
These phases affect the masses of the sparticles and the Higgs bosons
as well as their couplings to the SM particles and to each other.
Thus their presence can affect the phenomenology of the sfermions,
charginos/neutralinos and that of the Higgs bosons at the colliders. 
These phases can thus change even the CP-even variables such as the 
sparticle production rates, their decay widths and branching ratios.  Of course 
the 'direct' measure of these phases will be the non-zero value of  CP-odd
observables constructed out of the momenta of the final state decay products.

Effects of nonzero \cp\ phases on the search and study of $\ch, \nt$, sfermions
and the charged Higgses have been investigated in great detail~\cite{thsabro}. 
Due to the high precision of the measurements that would be possible at the 
ILC~\cite{ilc-ref},
at times the CP-even variables like the branching ratios, cross-sections, 
polarisations of fermions in final state, will offer a better probe of the \cp\
phases than the CP-odd quantities constructed out of the final state momenta.

For the hadronic colliders the effects of CPV in MSSM on Higgs 
phenomenology  have been studied in the context of Tevatron and the 
LHC~\cite{dedes,kane,extrahiggs,hagiwara,carena,francesca,charged,ggr1,ggr2}, 
whereas that on the  
$\ch, \nt$ phenomenology has been studied mainly only in the context of the 
Tevatron~\cite{choi1}. I will begin with a brief discussion 
of the latter  in the next subsection.
\subsection{Effect of \cp\ on $\ti\chi$ phenomenology at hadronic colliders}
The end point of the invariant mass distribution of the dilepton pair produced 
in the three body decay $\nt_2 \to \nt_1 l^+ l^-$, plays a very important 
role in SUSY phenomenology, particularly at the hadron colliders.  Nonzero
CPV phases can change this distribution substantially and thus can 
affect phenomenology of cascade decays of the sparticles. Further, this 
also opens up possibility of extracting  information on the CPV phase 
if {\it all} the SUSY model parameters are known. At a $p \bar p$ collider,
it is possible to construct CP-odd quantities, for $\tilde\chi^\pm, 
\tilde\chi^0$  system, producing a trilepton signal via $p \bar p \to
\ch_1 \nt_2 \to l l' \bar {l'} + \etm$. Since the initial state is a 
CP-eigenstate, it is possible to construct T-odd variables using the
initial (anti)proton direction. 
\begin{figure}
\centerline{
\includegraphics*[scale=0.4]{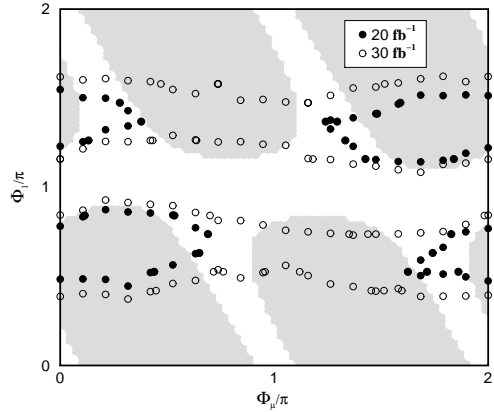}}
\caption{Values of the \cp\ phases $\Phi_\mu, \Phi_1$ that can be probed
using the CP/T-violating asymmetries for the trilpeton signal,
at $5 \sigma $ level, for the luminosity indicated on the 
plot\protect\cite{choi1}. Shaded regions are ruled out by the EDM constraints.}
\label{mono-tev}
\end{figure} 
Some of these variables are: 
$ {\cal O}_T=\vec{p}_{_{\ell_1}}\cdot (\vec{p}_{_{\ell_3}}
\times\vec{p}_{_{\ell_4}})$  and  
${\cal O}^{\ell\ell'}_T=\vec{p}_p\cdot\left(\vec{p}_\ell\times
          \vec{p}_{\ell'}\right).$
In the first case $\ell_1=\ell^-$ coming from the chargino decay 
$\tilde{\chi}^-_1\rightarrow
\tilde{\chi}^0_1\,\ell^-\bar{\nu}_\ell$, and $\ell_3=\ell^{\prime -}$,
$\ell_4=\ell^{\prime +}$ coming from the neutralino decay $\tilde{\chi}^0_2
\rightarrow \tilde{\chi}^0_1\,\ell^{\prime -}\ell^{\prime +}$.  
In the case of the second variable, the $\{\ell,\ell'\}$  stand for 
any combination of the two momenta among the three final state leptons.
The \cp\ phases also of course affect $\sigma(p\bar{p}
\rightarrow\tilde{\chi}^-_1\tilde{\chi}^0_2)$,
${\cal B}(\tilde{\chi}^-_1\rightarrow\tilde{\chi}^0_1\ell^-\nu)$ and 
${\cal B}(\tilde{\chi}^0_2\rightarrow\tilde{\chi}^0_1\ell^+\ell^-)$.
Fig.~\ref{mono-tev} illustrates what could be achieved at the Tevatron if an
integrated Luminosity $\L = 20 (30)$  \fb  were to be available.

Since LHC is a pp collider the initial state is not a CP eigenstate and hence
a different set of variables needs to be constructed in order to probe the \cp\
phases in the $\ti\chi$ studies at the LHC. Even more importantly, 
investigations into effects of these phases on the studies at the LHC using,
say, the cascade decays, are needed and do not yet exist. Such studies are
essential in order to assess the feasibility of determination of the SUSY
parameters at the LHC from sparticle phenomenology, which involve sparticle 
mass measurements using end point of the dilepton invariant mass spectrum.
\subsection{MSSM with CPV and Higgs phenomenology}
CP violation in the Higgs-sector is possible only in the presence of multiple
Higgs doublets, the simplest one being the two Higgs doublet model (2HDM).
In the CP-conserving 2HDM there exist three neutral Higgs boson states: the 
CP-even $h,H$ and CP-odd A. In presence of CP violation all these three mix and
one has three states $\phi_1,\phi_2,\phi_3$, none of which have a fixed 
CP property. Discussions of CP violation in the Higgs sector and hence of 
CP mixing among the neutral higgs boson states in a {\it model independent way},
existed in literature~\cite{old-cpvhiggs}. Effect of this mixing on the 
couplings of  the mixed CP states $\phi_1,\phi_2,\phi_3$ with a pair of gauge 
bosons/fermions i.e., $\phi_i f \bar f$, $\phi_i V V$, can change the Higgs 
phenomenology profoundly. It can be shown that various sum rules exist for 
these and we have for example,
$$
g_{\phi_i WW}^2 + g_{\phi_j WW}^2 + g_{\phi_k WW}^2 = g^2 m_W^2, i \neq j 
\neq k.
$$
As mentioned before the \cp\ phases of SUSY(breaking) parameters induce \cp\
violation in the Higgs sector, through loop corrections involving the third 
generation sfermions, even though the tree level scalar potential
conserves CP. Thus a CPV MSSM is distinguished from a general CPV 2HDM, by the
fact that the former has a prediction for the mixing in terms of SUSY(breaking)
\cp\ phases of the MSSM mentioned before.

\begin{figure}
\centerline{
\includegraphics*[scale=0.4, angle=90]{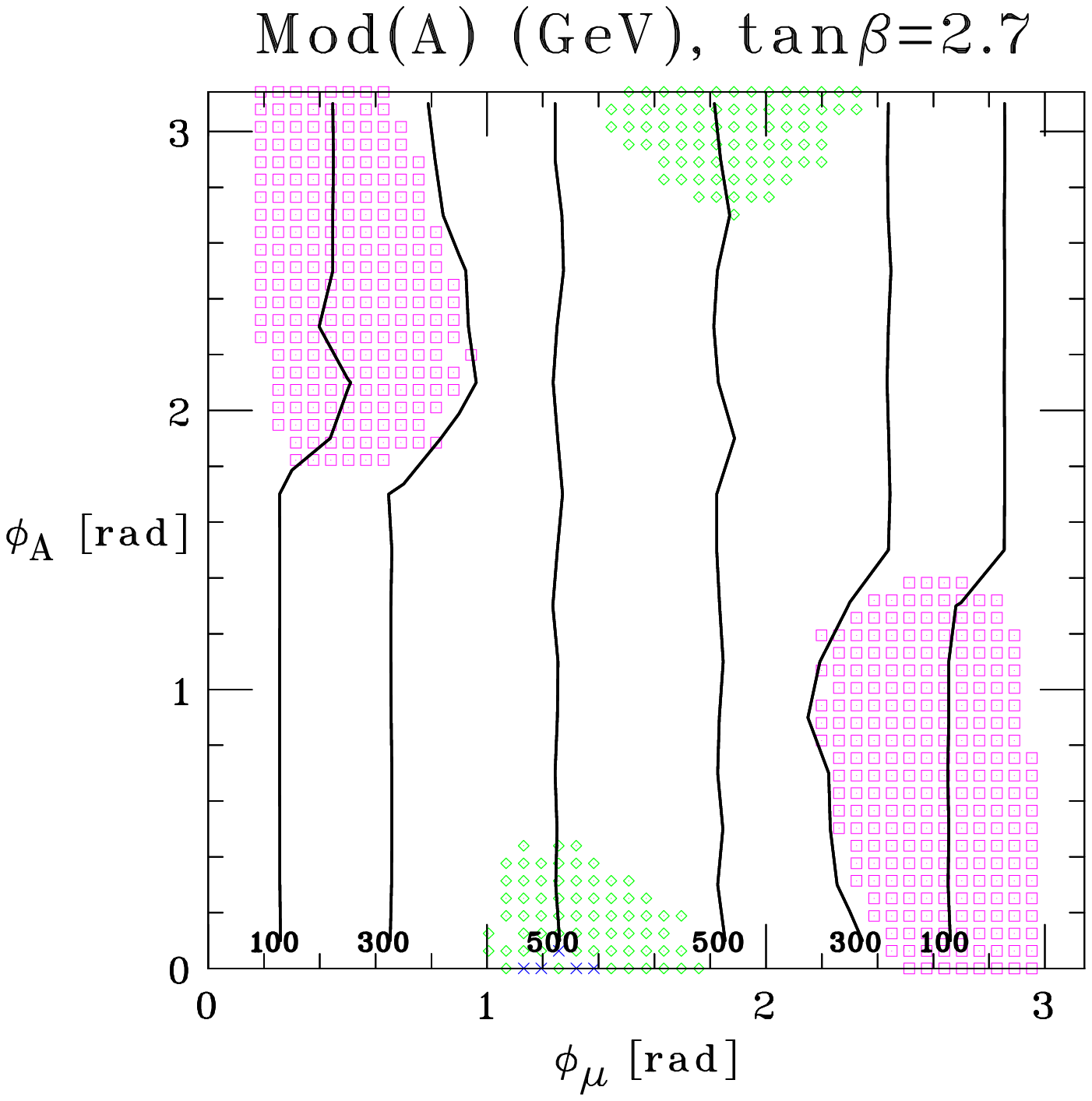}
\includegraphics*[scale=0.4, angle=90]{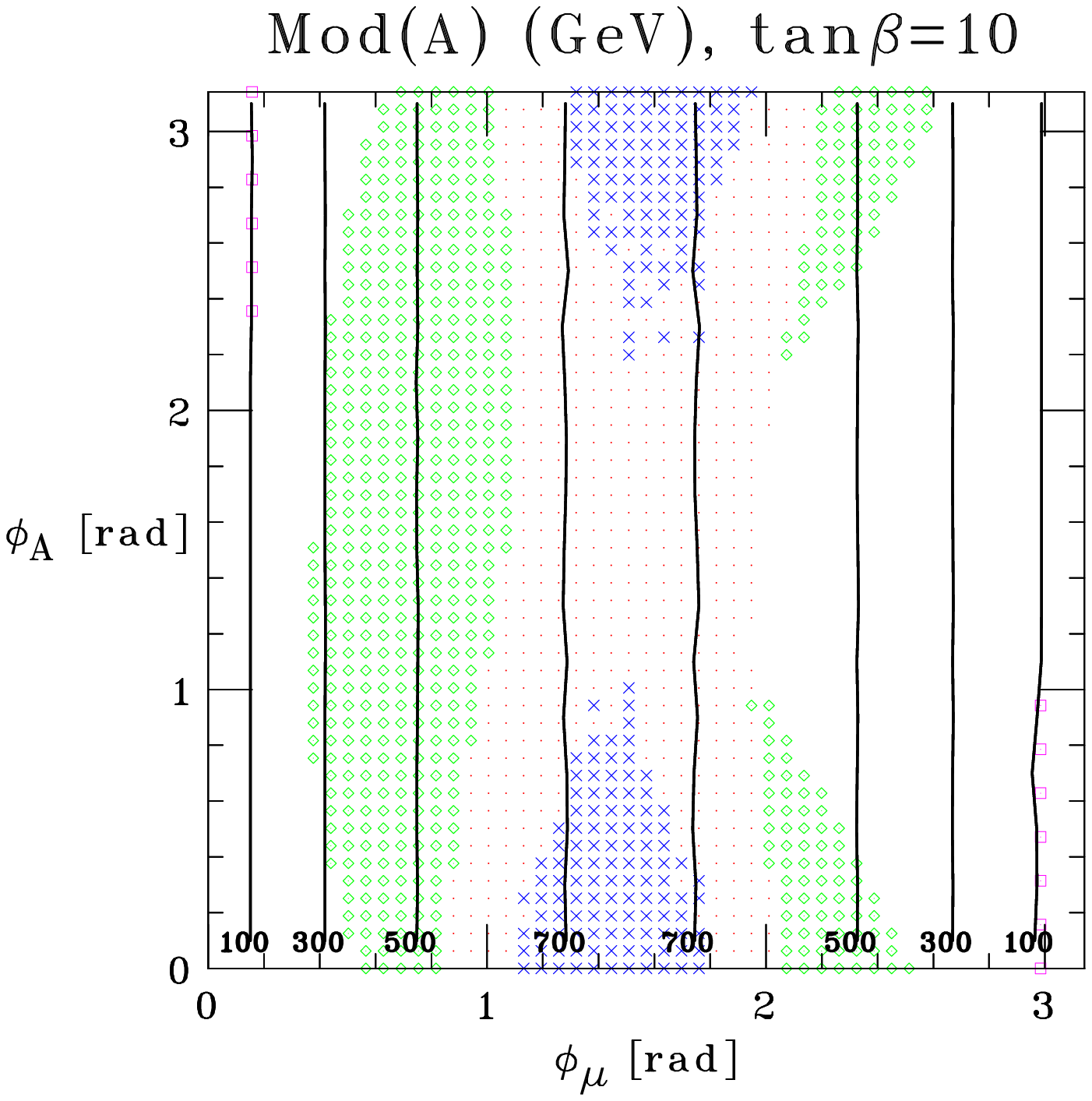}
}
\caption{Contours of $|A|$ in GeV in the $\Phi_\mu$--$\Phi_A$ plane along with
the regions excluded by constraints on the EDM's indicated by the shaded area 
~\protect\cite{dedes}. The plot in the left panel is  for $\tan \beta =2.7, 
|\mu| = 600 \protect\rm GeV, M_{\tilde q_{1,2}}=1000 {\rm GeV}, 
M_{\tilde q_{3}} = 300 {\rm GeV}, M_{\tilde g} = 300 {\rm GeV}$ and 
$ M_{A} = 200 {\rm GeV}$, whereas for the right panel $\tan\beta = 10$,
$M_{\tilde q_{1,2}}=300 {\rm GeV}.$}
\label{edm}
\end{figure}

In general the scalar potential for a 2HDM can be written as 
\begin{eqnarray}
V &=& m_{11}^2 \Phi_1^{\dagger} \Phi_1
+ m_{22}^2 \Phi_2^{\dagger} \Phi_2
- [ m_{12}^2 \Phi_1^{\dagger} \Phi_2 + h.c.] \nonumber \\
&&+\half \lambda_1 (\Phi_1^{\dagger} \Phi_1)^2
+\half \lambda_2 (\Phi_2^{\dagger} \Phi_2)^2
+\lambda_3 (\Phi_1^{\dagger} \Phi_1)(\Phi_2^{\dagger} \Phi_2)
+\lambda_4 (\Phi_1^{\dagger} \Phi_2)(\Phi_2^{\dagger} \Phi_1) \nonumber \\
&&
+\left\{ \half  \lambda_5 (\Phi_1^{\dagger} \Phi_2)^2
+ \left[ \lambda_6 (\Phi_1^{\dagger} \Phi_1)
+ \lambda_7 (\Phi_2^{\dagger} \Phi_2) \right] \Phi_1^{\dagger} \Phi_2
+ h.c. \right\} \nonumber
\end{eqnarray}
Unitarity implies that $V \in  \Re $, which in turn means that
$ \{m_{11},m_{22},\lambda_{1-4}\} \in {\bf \Re } $ and 
$\{m_{12},\lambda_{5-7}\} \in {\bf \cal C}$.  In the MSSM various parameters
in this potential can be  expressed in terms of the gauge couplings and 
SUSY parameters $\mu, B$, with $\lambda_5=\lambda_6=\lambda_7=0$. It can be 
shown in this case that any nonzero phase that $m_{12}^2$ may have can be 
rotated away by a redefinition of fields. 
Thus the tree level MSSM higgs potential can be CP-conserving even with 
nonzero phase of $\mu$. However, at loop level, diagrams such as shown in 
\begin{figure}
\centerline{
\includegraphics*[scale=1.1]{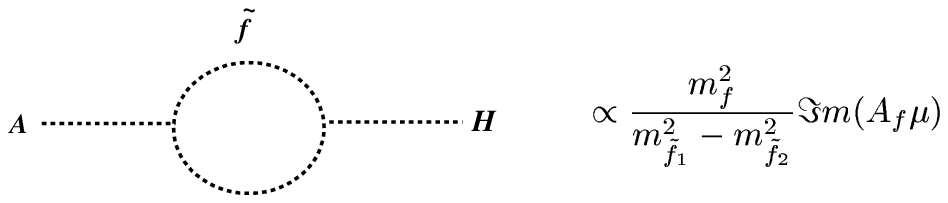}}
\caption{Loop diagrams inducing CP-mixing in Higgs sector in the \cp\ MSSM}
\label{mixing_dia}
\end{figure}
Fig.~\ref{mixing_dia} 
give non vanishing complex contribution to $m_{12}^2$ which can not be any more
rotated away as there is no freedom of field redefinition anymore.
The CP-mixing in the Higgs sector can  be parametrised by 
$\{\Phi_{A_f},\Phi_3,\Phi_{\mu} \}$~\cite{apostolos,cpx}. 
In certain regions of SUSY(breaking) parameter space the \cp\ phases can 
also induce CP violation in the sfermion-sfermion-Higgs vertex~\cite{dedes}
and this in turn can give rise to EDM's of fermions which depends on
$|A|, \Phi_\mu$ and $\Phi_A$. Fig.~\ref{edm} shows the constraints on the
phases given by the EDM's for a given set of sparticle masses and SUSY 
parameters. Thus if this scenario is realised one will have to choose $|A|$
values to be greater than indicated on the contours so as to satisfy the
EDM constraints and look at the effect of this CP-mixing on the Higgs
boson phenomenology. The right panel of the figure shows that it is much 
more difficult to achieve consistency with the data  for larger values of 
$\tan \beta$. In this case the allowed regions are due to accidental SUSY 
cancellations.
For large values of $M_{\tilde q_{3}}$ and $A$, the loop induced CP-mixing 
mentioned earlier becomes dominant. In the so-called CPX scenario~\cite{cpx},
designed to showcase this mixing one chooses:
$M_{\tilde Q_3}=M_{\tilde U_3}=M_{\tilde D_3}=M_{\tilde L_3}=M_{\tilde E_3}
=M_{\rm SUSY}, \mu = 4 M_{\rm SUSY}$, $|A_{t,b,\tau}|=2 M_{\rm SUSY}$ and 
$|M_3| = 1 TeV$. Then the masses and couplings of the Higss bosons are studied 
as functions of $\tan \beta, M_{H^\pm}, \Phi_{A_f}, \Phi_\mu, \Phi_3$ as well as
the  SUSY scale $M_{\rm SUSY}$. In this case 
the EDM constraints are  easily  satisfied for the chosen parameters and 
hence the phases $\Phi$ can be varied freely. For obvious reasons the 
phases $\Phi_{A_t}, \Phi_{A_b}$ dominantly  affect the masses and couplings 
of the mixed Higgs boson states.  
\begin{figure}
\centerline{
\includegraphics*[scale=0.5]{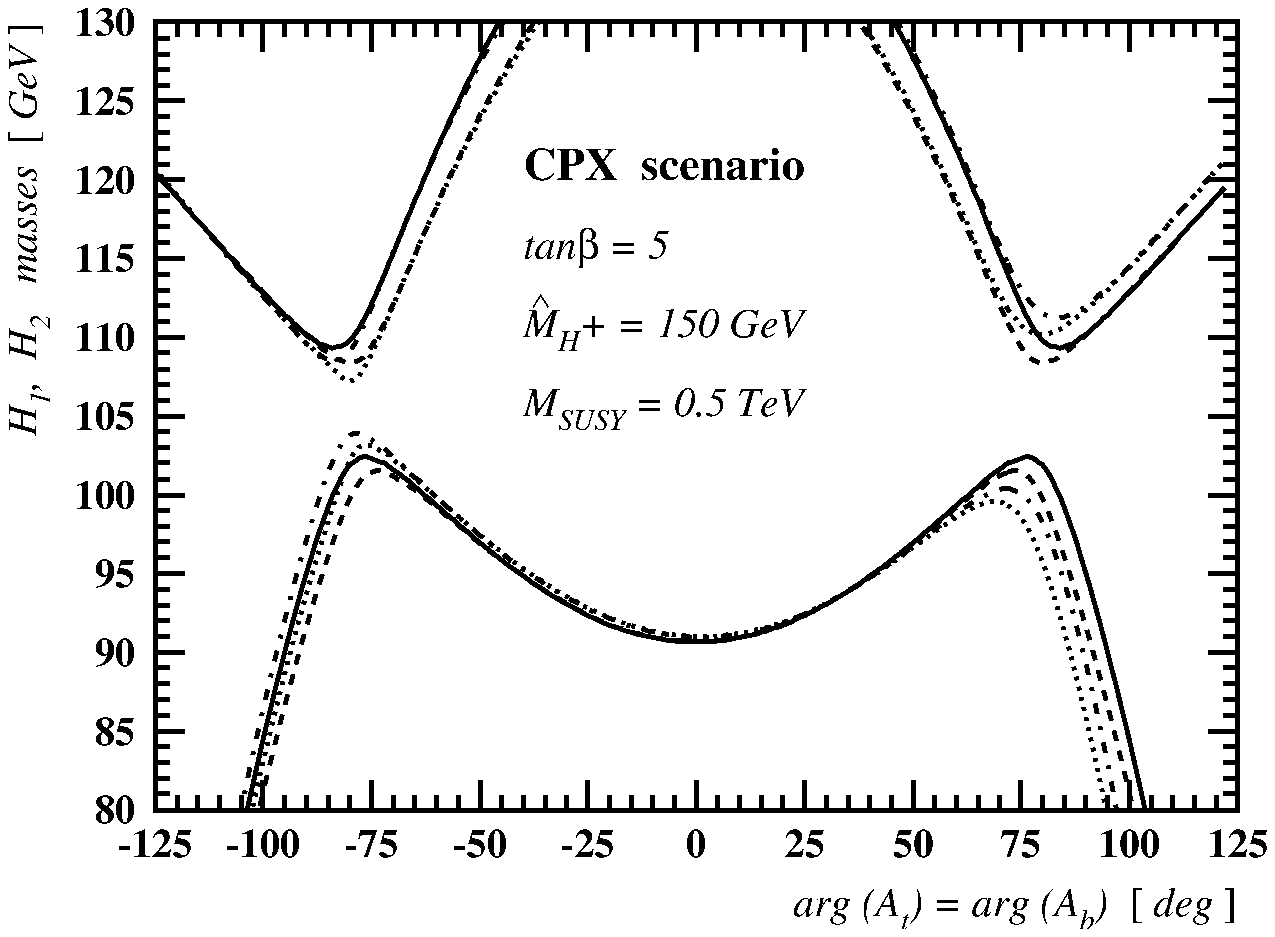}
\includegraphics*[scale=0.5]{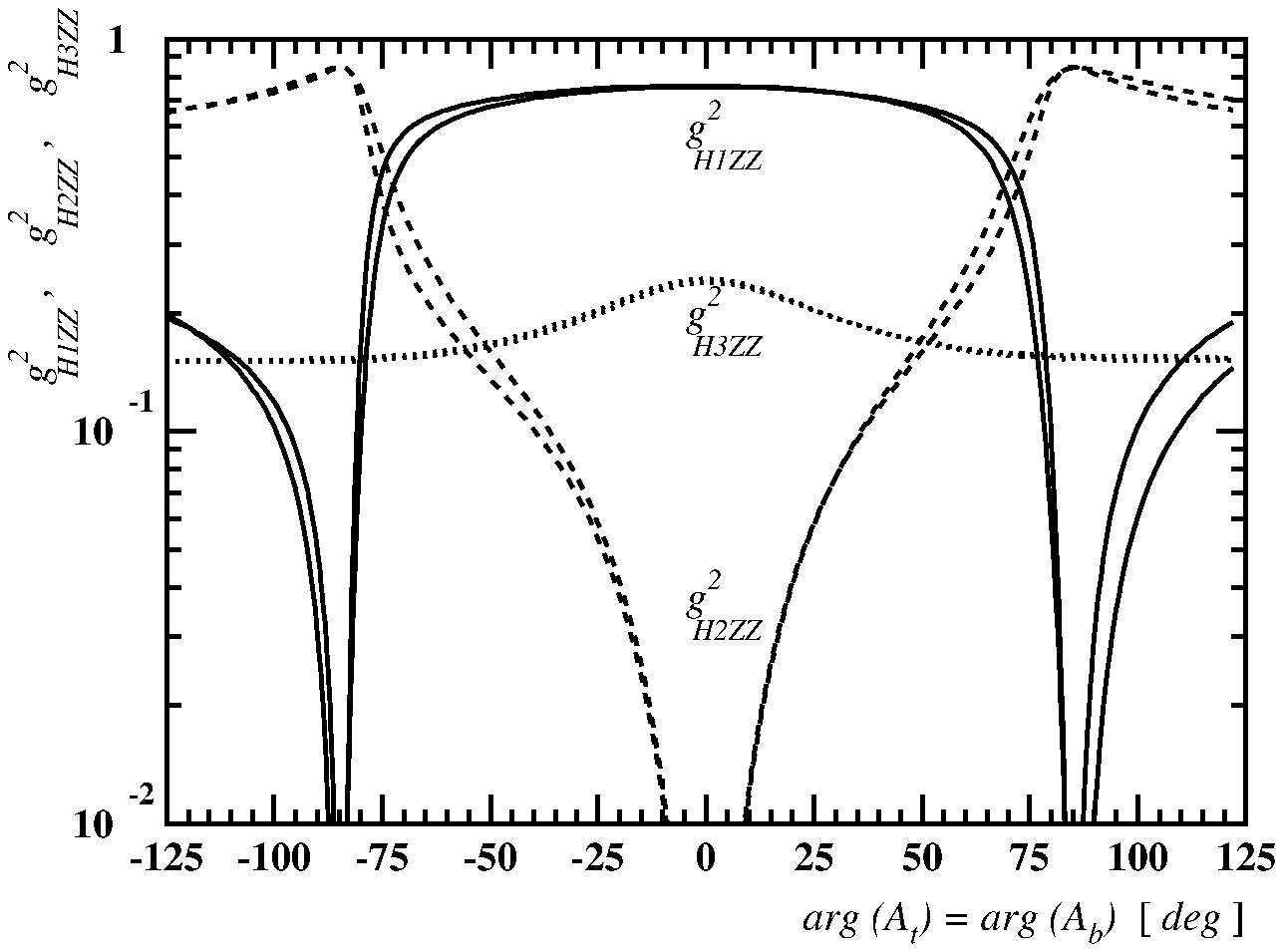}
}
\caption{Variation of $M_{\phi_i}$ and $g^2_{\phi_i VV}$ with 
$\Phi_{A_t} = \phi_{A_b}$. $\Phi_\mu = 0$ and $\Phi_3 = 0 (\pi/2)$.
Values of all the other relevant parameters are  indicated on the figure  
and correspond to the case where the $\phi_3$ is also light with a mass
$\sim 150$ GeV~\protect\cite{carena}.}
\label{mixing}
\end{figure}
The left panel of the Fig.~\ref{mixing} shows the masses for the lightest two 
Higgs-boson states and the right panel shows its couplings to a pair of vector 
boson V. It is to be noted that the $\Phi_3$ has an effect on the masses 
at two loop level and hence the dependence on it is quite weak for
small values of $\tan \beta$.  Further it is also seen that for large phases 
of $A_t, A_b$, $g_{\phi_{1}ZZ}$ decreases and it can even vanish for the case 
where $\phi_1$ is mostly a pseudoscalar.  For larger values of $\tan\beta$
effects of the phase $\Phi_3$ can be significant and have been investigated 
in Ref.~\cite{francesca}.

Since the production of Higgs boson at all the colliders utilises its large 
couplings with the $Z/W$ bosons as well the heavy fermions, it is clear that 
the above change in the couplings can affect the Higgs boson phenomenology at 
all the colliders drastically. The non observation of a Higgs boson signal in the
direct searches at the LEP needs to be reinterpreted in the MSSM with CP 
violation. The recent analysis from OPAL~\cite{opallimit} shows that indeed 
there are 'holes' in the excluded region at small $\tan \beta$ and $m_{\phi_1}$ 
in the $\tan \beta$--$m_{\phi_1}$ plane that are allowed  with 
the non-observations of the signal at LEP. Essentially the lightest mass
eigenstate is dominantly a pseduoscalar in this case and hence does not couple
to a $ZZ$ pair very effectively.  
\begin{figure}
\centerline{
\includegraphics*[scale=0.6]{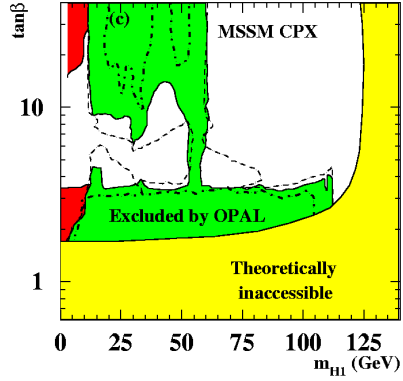}
}
\caption{Regions in the $\tan \beta$--$m_{\phi_1}$ plane disallowed 
theoretically or excluded  by the current LEP searches\protect\cite{opallimit}.
The allowed `hole' at the low $\mh{+}, \tan\beta$ values can be seen very 
clearly.}
\label{opalexclusion}
\end{figure}

\subsection{Effect of CP mixing on Higgs searches at Hardonic Collider}
At the Tevatron and at the LHC gluon fusion provides the main production mode
for the Higgs. The loop induced $gg \phi_i$ coupling is dominated by the $t, 
\ti t$ and $\ti b$ loops. CP violation in the MSSM can have effects on this 
loop induced coupling and  thus affect the Higgs production rates at the 
hadronic colliders.
\begin{figure}
\centerline{
\includegraphics[angle=90,width=6cm,clip=true]{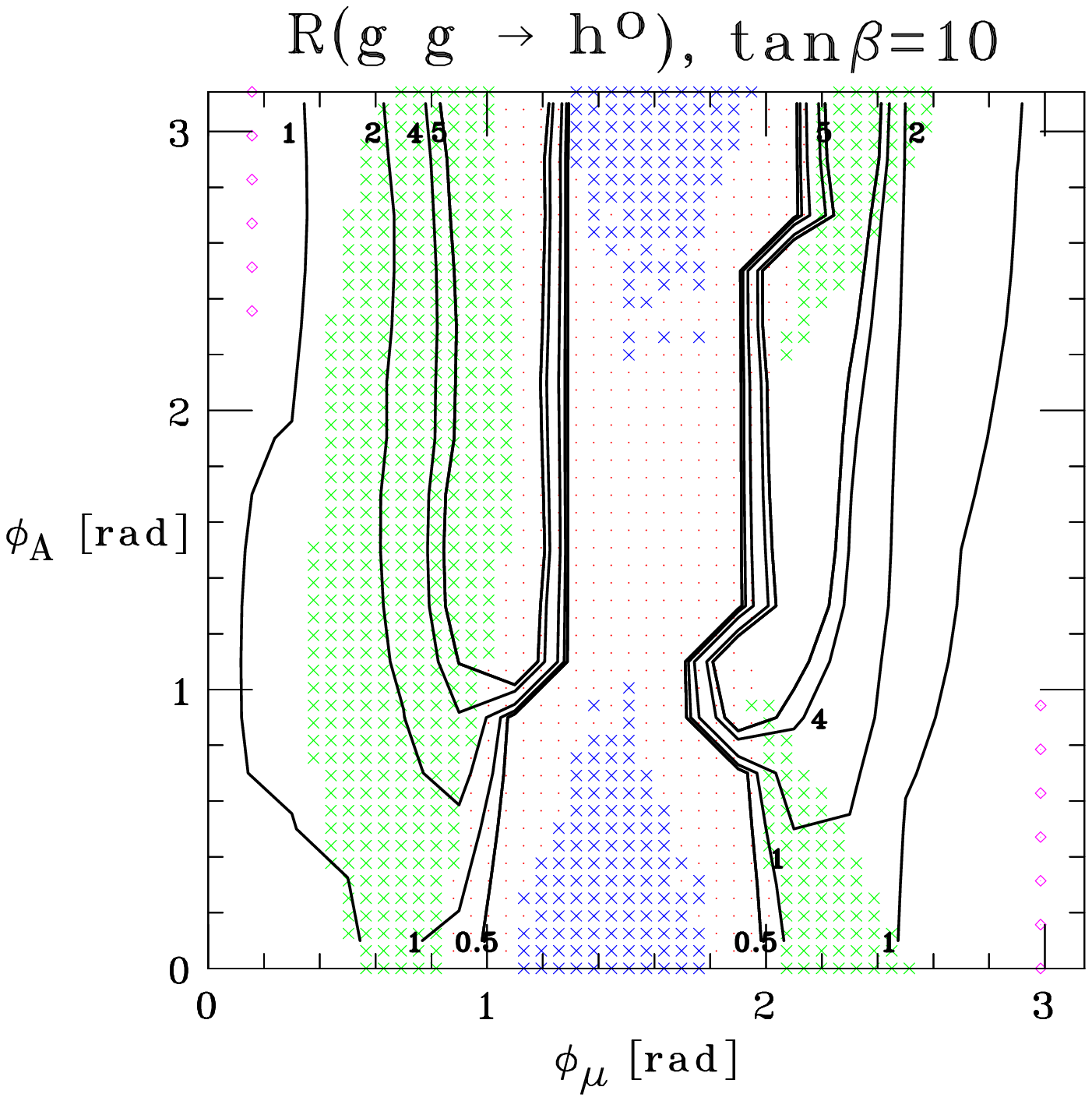}
\includegraphics[angle=90,width=6cm,clip=true]{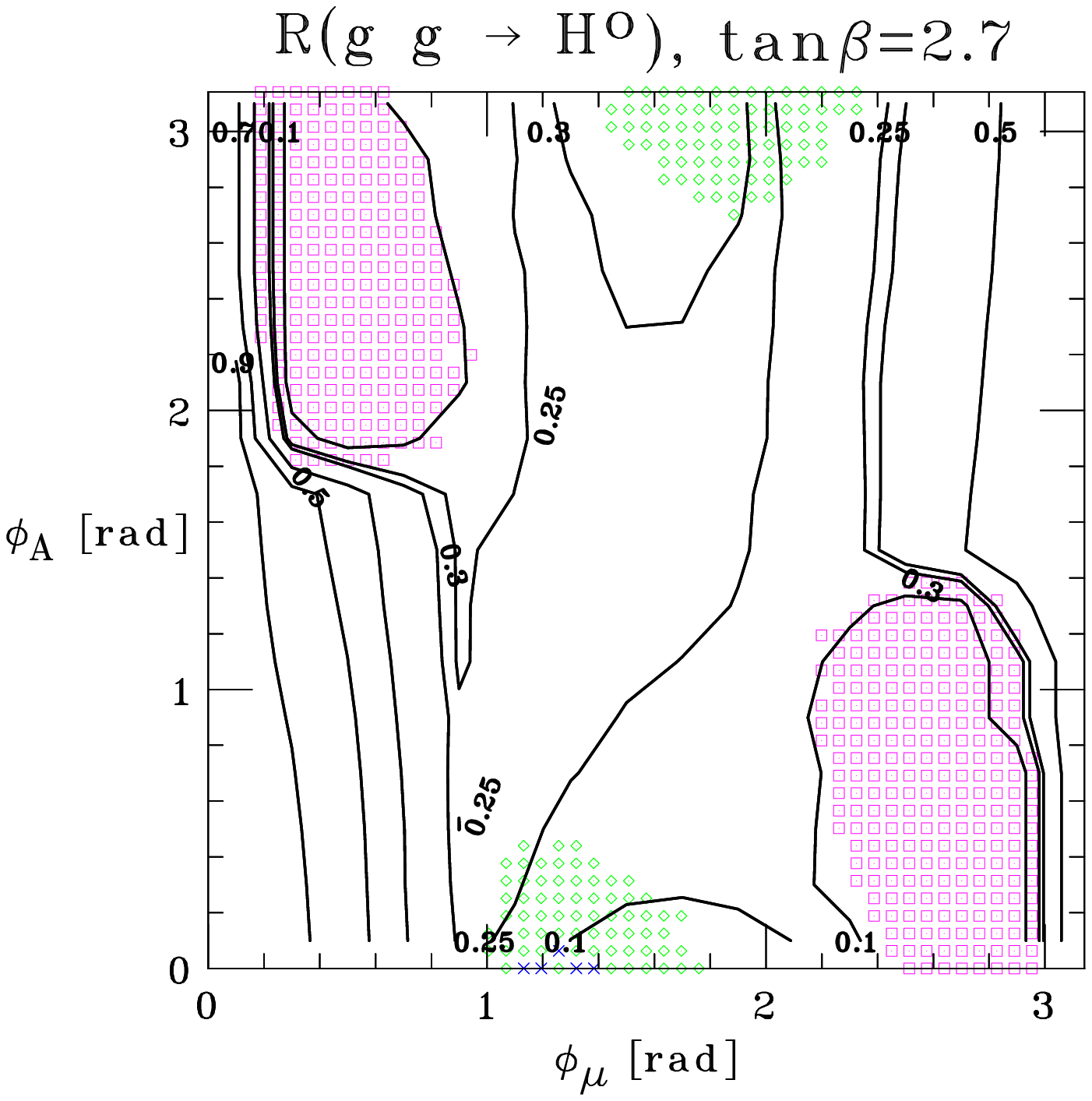}
}
\caption{Contours of ratio of Higgs production to that expected in the 
CP conserving case, as a function of $\Phi_\mu$ and 
$\Phi_A$\protect\cite{dedes}. The left panel is for $h$ and $\tan \beta =10$
and the right panel is for $H$ and for $\tan \beta = 2.7$. Also shown 
are the regions disallowed by the EDM constraints.}
\label{dedeshiggs} 
\end{figure}
In Fig.~\ref{dedeshiggs} the contours of ratios of  $h,H$ production rates in 
the CP violating MSSM to those without CP violation are shown. This
corresponds to the  case  where the \cp\ in the MSSM
induces CPV $\ti q \ti q h (H)$ couplings. As expected from the sum rule
we find that whereas the $h$ production rate increase in the allowed region, 
the $H$ production rate  decreases.

In case of loop induced CP mixing in the Higgs 
sector~\cite{apostolos,cpx} a complete analysis involving all the three
colliders LEP,Tevatron and LHC was performed~\cite{carena}. 
\begin{figure}
\centerline{
\includegraphics[scale=0.4]{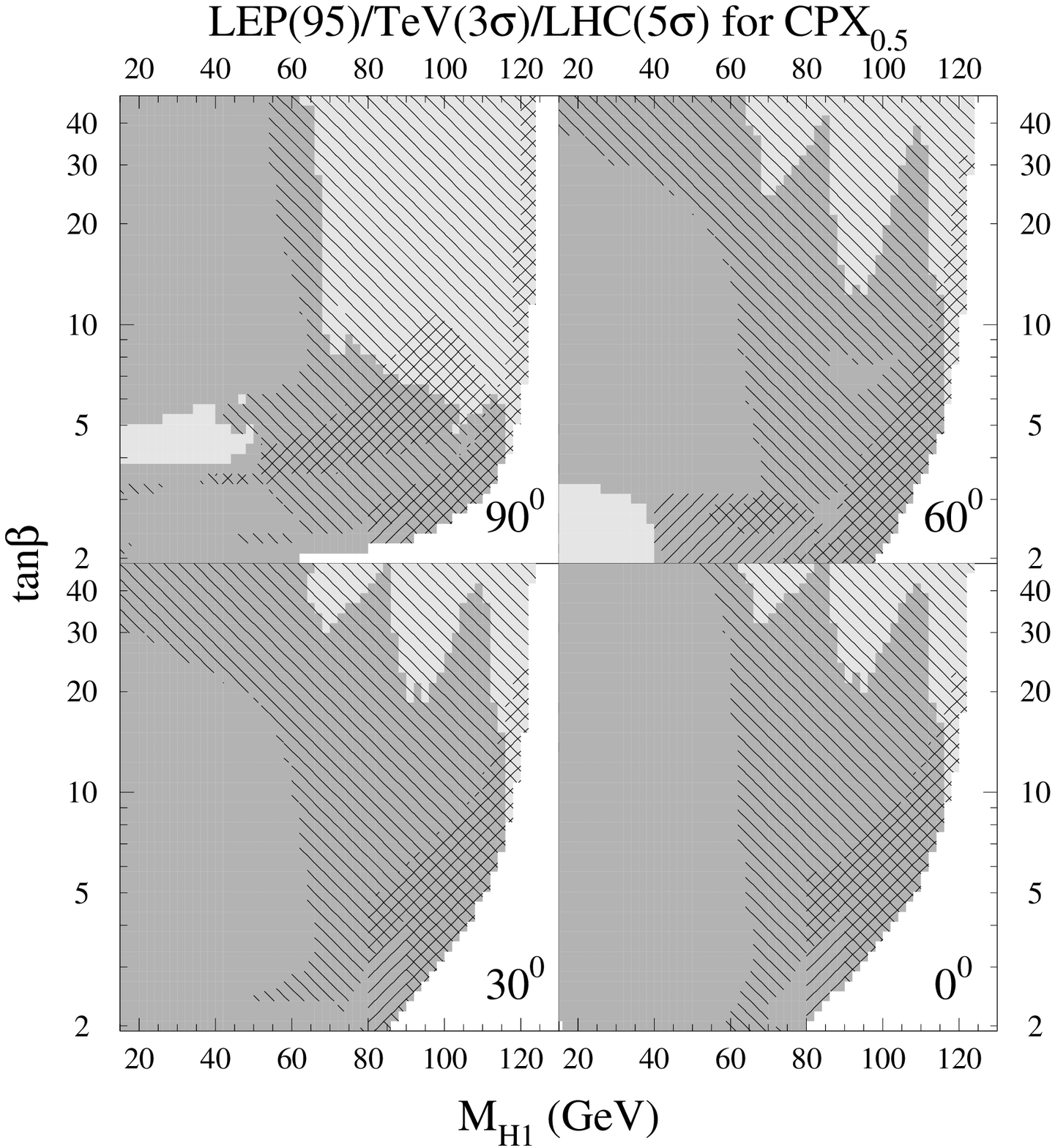}
\includegraphics[scale=0.4]{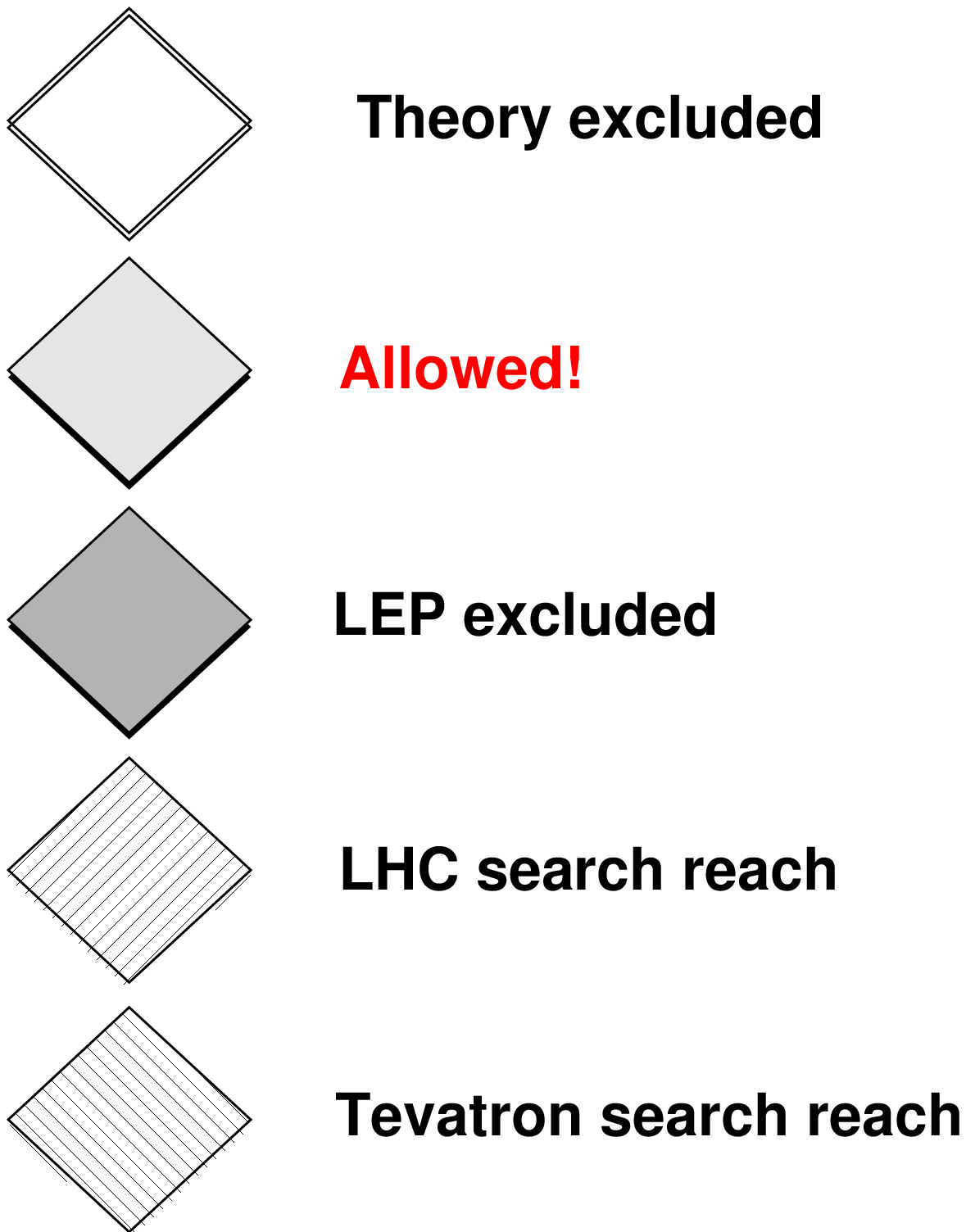}}
\caption{Coverage of LEP,Tevaron and the LHC for the Higgs searches
in CPX scenario\protect\cite{carena}}
\label{gaps}
\end{figure}
In addition to the gaps in the LEP coverage for small Higgs masses which
is already evident in the OPAL results of Fig~\ref{opalexclusion}, this figure 
also shows that neither the Tevatron nor the LHC have reach in the same region
due to a reduced $t \bar t \phi$ coupling, along with a reduction in 
$V V \phi$ coupling there. Thus the issue of light higgs searches at the LHC 
needs to be revisited for the CP violating MSSM. Preliminary analyses by ATLAS
collaboration~\cite{shumacher} seems to confirm this result of the theory 
analysis.

\subsection{Search for a light $\phi_1$ in $H^\pm$ decay at the LHC}
\begin{figure}
\centerline{
\includegraphics*[scale=0.21] {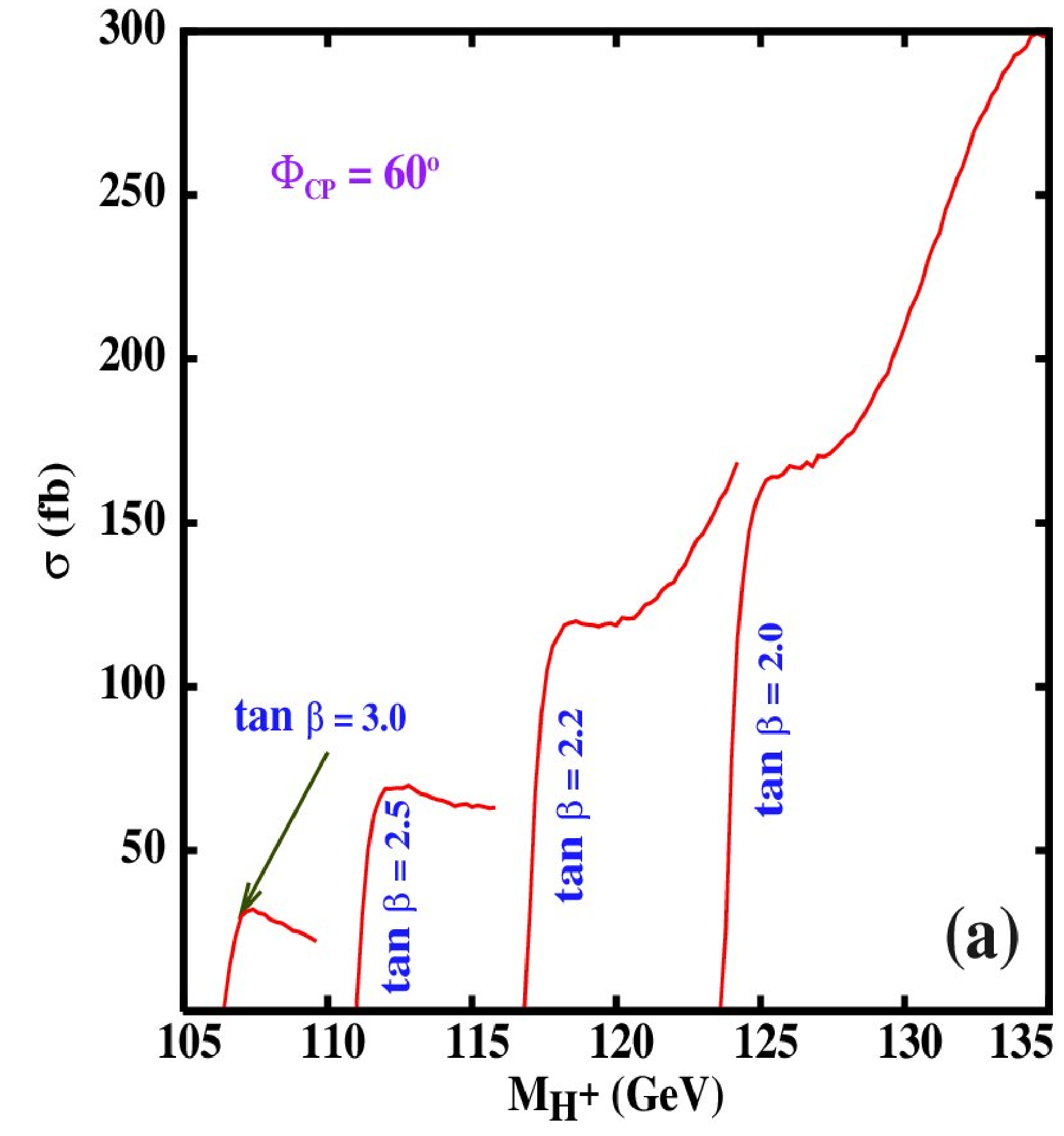}
\hskip 0.5cm
\includegraphics*[scale=0.30]{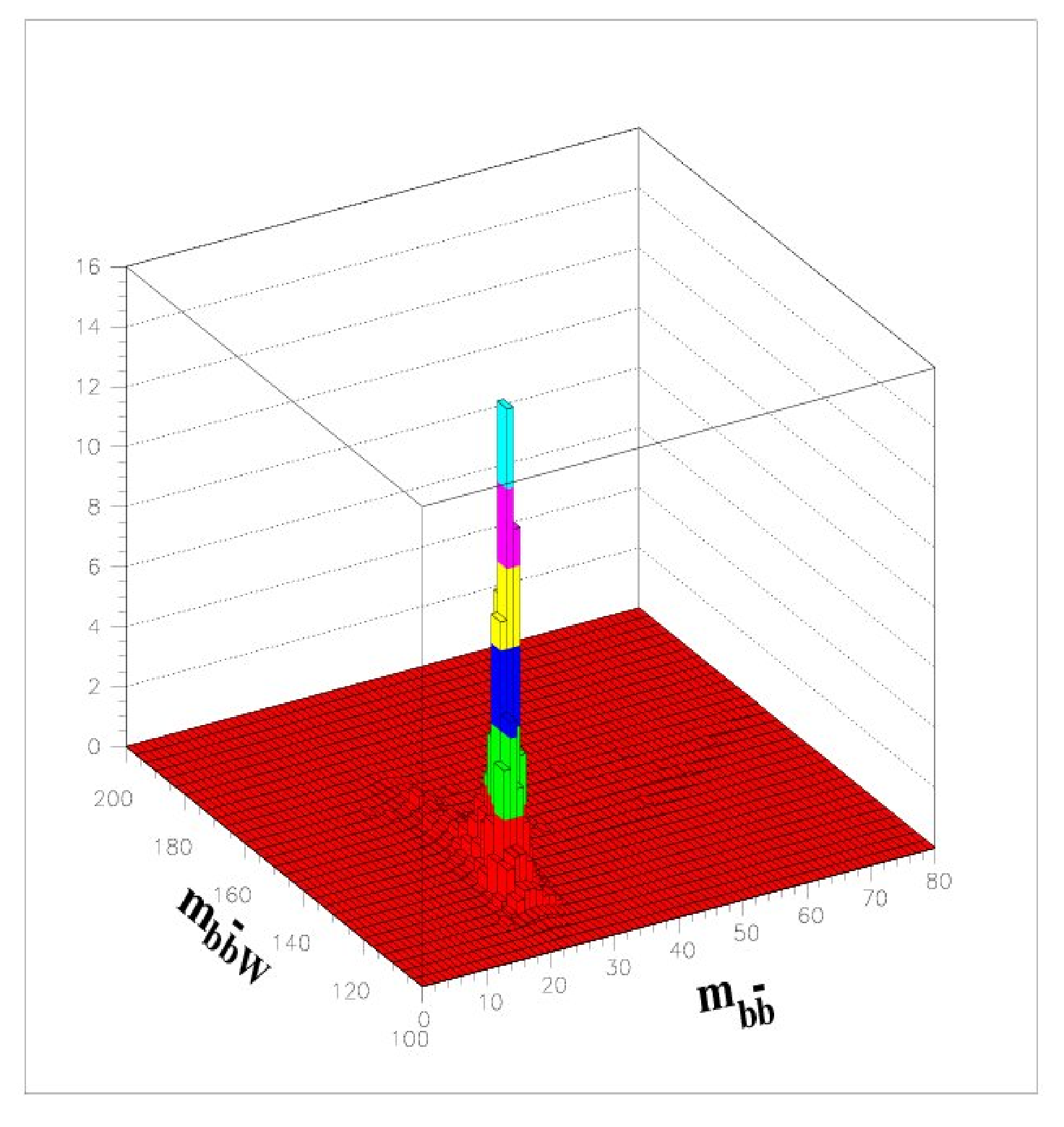}}
\caption{Left panel shows variation of the expected  cross-section with 
$M_{H^+}$  for four values of $\tan\beta =2, 2.2, 2.5 $ and $3$.
The CP-violating phase $\Phi_{\rm CP}$ is $60^{\circ}$. The right panel 
shows lustering of the $b\bar b, b\bar bW $ and $b\bar b b W$ invariant
masses in the three-dimensional plot for the correlation between
$m_{b\bar b}\equiv M_{H_1}$ and $m_{b\bar b W}\equiv M_{H^+} $ invariant 
mass distribution. Details of the parameters used are given in~\protect\cite{ggr2}.}
\label{ggrfig}
\end{figure}

One possible way this 'hole' could be probed is by searching for a light
$\phi_1$ in the decay of the charged Higgs $H^\pm$~\cite{ggr1,ggr2}. 
The parameter space  where the hole occurs corresponds to a relatively light
$H^\pm$ ($M_{H^\pm} < M_t$) , which is predicted to decay dominantly into
the $W \phi_1$
channel. Thus one expects to see a striking $t \bar t$ signal at the LHC,
where one of the top quarks decays into the $bb \bar b W$ channel, via
$t \to b H^\pm, H^\pm \to W \phi_1$ and $\phi_1 \to b \bar b$. The 
characteristic
correlation between the $b \bar b$, $b \bar b W$ and $b b \bar b W$ invariant
mass peaks is expected to make this signal practically free of the SM
background. Our parton level Monte Carlo simulation yields up to 4500 events,
for ${\cal L} = 30$ fb$^{-1}$, over the parameter space of interest, after
taking into account the b-tagging efficiency for three or more b-tagged jets.
The clustering of the invariant mass of the $b \bar b$ pair with the smallest 
value around $m_{\phi_1}$ and that of the  $b \bar b  W$ invariant mass 
around $M_{H^+}$ can be seen from Fig~\ref{ggrfig} taken from 
Ref.~\cite{ggr2}. This result needs to be confirmed by experimental simulations.
including detector effects.
\section{Conclusion}
Thus we note that the possibilities of probing the CP-violating phases in the 
MSSM in sparticle production and decays at the LHC have yet to be explored 
fully. These CPV phases can, in principle, affect the shape of dilepton
invariant mass spectrum for the dilepton pair produced in the decay of $\nt_2$
and thus affect the sparticle mass determination accuracy etc. Further these
modifications may be a probe of the CPV phases if remaining SUSY parameters 
are known. CP conserving quantities such as cross-sections, branching ratios
are sensitive to the CPV phases, but for direct measurements CPV variables 
need be constructed. This task has still to be done for the LHC.

CP violation in MSSM can affect the Higgs search possibilities at the
LEP and LHC profoundly. For low $m_A$ and not too heavy squarks, \cp\
MSSM parameters can induce CPV in the $\ti q \ti q \phi$ vertex, which in 
turn can affect the Higgs production rate through gluon fusion, by as much
as a factor 10, for values of CPV phases which are consistent with the EDM
constraints. In the CPX scenario~\cite{cpx} chosen to showcase the \cp\
in the MSSM, existence of a light neutral Higgs boson ($M_{\phi_1} \lts 50$ 
GeV) is allowed low $\tan \beta (\lts 5)$ region and could have escaped the 
LEP searches 
due to a strongly suppressed $\phi_1 Z Z$  coupling. Even the LHC might miss
discovering such a $\phi_1$ due to the suppression of the $t \bar t \phi_1$
coupling as well. In this situation, decay of the light $H^\pm \to \phi_1 W$
may provide a signal for the $\phi_1$ through its $b \bar b$ decay.
Thus one expects to see a striking $t \bar t$ signal at the LHC,
where one of the top quarks decays into the $bb \bar b W$ channel, via
$t \to b H^\pm, H^\pm \to W \phi_1$ and $\phi_1 \to b \bar b$.

\bigskip
Acknowledgment:     
{\small 
I would like to thank the organizers for an enjoyable and  interesting  meeting 
in this wonderful city. I would like to acknowledge, thankfully, the partial 
support I received from the Austrian Academy of Sciences towards attending the 
meeting. Thanks are due to J.S. Lee for suggestions and comments on the
first version of this article.}


\bigskip


\begin{thebibliography}{99}

\bibitem{bellebabar}
BELLE Collab., A. Abashian et al.,
    Phys. Rev. Lett. {\bf 86}, 2509 (2001), hep-ex/0102018;
BABAR Collab., B. Aubert et al.,
    Phys. Rev. Lett. {\bf 86}, 2515 (2001), hep--ex/0102030.

\bibitem{book} For an account of  SUSY in High Energy Physics, see for example,
'Theory and Phenomenology of Sparticles', M. Drees, R. M. Godbole and P. Roy, Ed. World Scientific, Spring 2004.

\bibitem{nath}  For some early references, see for example,
T.~Ibrahim and P.~Nath, Phys.\ Lett.\ B {\bf 418}, 98 (1998) [hep-ph/9707409];
M.~Brhlik, G.~J.~Good and G.~L.~Kane, Phys.\ Rev.\ D {\bf 59}, 115004 (1999) 
[hep-ph/9810457], A.~Bartl, T.~Gajdosik, W.~Porod, P.~Stockinger 
and H.~Stremnitzer, Phys.\ Rev.\ D {\bf 60}, 073003 (1999) [hep-ph/9903402],
T. Falk, K. A. Olive, M. Pospelov and R. Roiban, \ Nucl. \ Phys.  B {\bf 560}, 
3 (1999), [arXiv:hep-ph/9904393].

\bibitem{thsabro}  
T.~Gajdosik, R.~M.~Godbole and S.~Kraml, JHEP {\bf 0409}, 051 (2004) 
[arXiv:hep-ph/0405167] and refernces therein.

\bibitem{ilc-ref}  
J.~A.~Aguilar-Saavedra {\it et al.}  [ECFA/DESY LC Physics Working Group
                  Collaboration], arXiv:hep-ph/0106315;\\
T.~Abe {\it et al.}  [American Linear Collider Working Group
Collaboration], in {\it Proc. of Snowmass 2001}, ed. N.~Graf,
arXiv:hep-ex/0106055;\\
K.~Abe {\it et al.}  [ACFA Linear Collider Working Group Coll.],
arXiv:hep-ph/0109166;
see: {\tt lcdev.kek.jp/RMdraft/}~.

\bibitem{dedes}  
A.~Dedes and S.~Moretti, \ Phys. \ Rev.  \ Lett.  {\bf 84}, 22 (2000),
[arXiv:hep-ph/9908516], \ Nucl. \ Phys. \ B {\bf 576}, 29  (2000), 
[arXiv:hep-ph/990941].

\bibitem{kane}
G.~L.~Kane and L.~T.~Wang, Phys.\ Lett.\ B {\bf 488}, 383 (2000)
  [arXiv:hep-ph/0003198].


\bibitem{extrahiggs}
A.~Arhrib, D.~K.~Ghosh and O.~C.~W.~Kong,
        Phys.\ Lett.\ B {\bf 537} (2002) 217 [hep-ph/0112039].

\bibitem{hagiwara}
S. Y. Choi, K. Hagiwara and J.S. Lee, Phys. \ Lett. \ B {\bf 529}, 212
(2002), [arXiv:hep-ph/0110138].

\bibitem{carena}
M.~Carena, J.~R.~Ellis, S.~Mrenna, A.~Pilaftsis and C.~E.~Wagner,
\ Nucl. \ Phys. \ B {\bf 659} 145 (2003), [arXiv:hep-ph/0211467].


\bibitem{francesca}
F.~Borzumati, J.~S.~Lee and W.~Y.~Song,
Phys.\ Lett.\ B {\bf 595}, 347 (2004) [arXiv:hep-ph/0401024].


\bibitem{charged}
E.~Christova, H.~Eberl, W.~Majerotto and S.~Kraml,
Nucl.\ Phys.\ B {\bf 639} (2002) 263,
Erratum-ibid.\ B {\bf 647} (2002) 359 [hep-ph/0205227];
E.~Christova, H.~Eberl, W.~Majerotto and S.~Kraml,
JHEP {\bf 0212} (2002) 021 [hep-ph/0211063].


\bibitem{ggr1}
N.~K.~Mondal {\it et al.}, Pramana {\bf 63}, 1331 (2004) [arXiv:hep-ph/0410340].

\bibitem{ggr2} D.~K.~Ghosh, R.~M.~Godbole and D.~P.~Roy, arXiv:hep-ph/0412193.


\bibitem{choi1} 
S.~Y.~Choi, M.~Guchait, H.~S.~Song and W.~Y.~Song, Phys.\ Lett.\ B {\bf 483}, 
168 (2000) [arXiv:hep-ph/9904276]; arXiv:hep-ph/0007276,


\bibitem{old-cpvhiggs} 
A.~Mendez and A.~Pomarol,  Phys. Lett. B {\bf 272} 313 (1991),
J.Gunion, H. Haber and J.~Wudka, Phys. Rev. D {\bf 43} 904 (1991),
B.Grzadkowski, J.Gunion and J.~Kalinowski, Phys. Rev. D {\bf 60} 075011
(1999).

\bibitem{apostolos}
A. Pilaftsis, \ Phys. \ Rev. D {\bf 58}, 096010 (1998),[arXiv:hep-ph/9803297]
and Phys. \ Lett. B {\bf 435}, 88 (1998), [arXiv:hep-ph/9805373],
S.~Y.~Choi, M.~Drees and J.~S.~Lee,
\ Phys. \ Lett. \ B {\bf 481}, 57 (2000), [arXiv:hep-ph/0002287],
M.~Carena, J.~R.~Ellis, A.~Pilaftsis and C.~E.~Wagner,
\ Nucl. \ Phys. \ B {\bf 586}, 92 (2000), [arXiv:hep-ph/0003180].

\bibitem{cpx} A.~Pilaftsis and C.~E.~Wagner, \ Nucl. \ Phys. \ B {\bf 553}, 
3 (1999), [arXiv:hep-ph/9902371].  

\bibitem{opallimit}  G.~Abbiendi {\it et al.} [OPAL Collaboration],
{\em Eur. Phys. J. C} {\bf 37} (2004).

\bibitem{shumacher} 
M. Schumacher, at  the meeting on 'CP violation and nonstandard Higgs'
//http://kraml.home.cern.ch/kraml/CPstudies/

\end{thebibliography}
\end{document}